\documentclass[twocolumn,showpacs]{revtex4}

\usepackage{graphicx}%
\usepackage{dcolumn}
\usepackage{amsmath}

\makeatother

\begin{document}

\preprint{HEP/123-qed}

\title[Short Title]{Interwoven basin structures 
of double logistic map \\
at the edge of chaos
}
\author{Gyu-Seung Shin}
\email{shings@khu.ac.kr}
\affiliation{College of Electronics and Information,
Institute of Natural Sciences, \\
Kyung Hee University, Yongin, Kyonggi 449-701, South Korea
}%

\date{\today}

\begin{abstract}

The carpeting regularity of
basin structures for simply
doubled logistic map is studied.
Examining the elementary structures
of which global basin structures
are composed, we found them to have a impressive 
interrelation 
which we call \emph{all in one and one in all}.
Curves appearing in the basins and 
their expressions are also discussed.
\end{abstract}
\pacs{05.45.-a}
\maketitle

We occasionally come across some patterns
where several geometrical structures are interwoven exquisitely.
Those structures seem to have very curious interrelations 
such that they are interwoven into others repeatedly
over a wide range of scale. 
Consequently,
a small section of the structure may contain 
information about the whole. 
Interwoven patterns and their regularity deserve to be
objects of aesthetical appreciation and scientific research.
Prominent instances would be those
patterns appearing in Mandelbrot sets and Julia sets \cite{j1}.
Also, dynamic systems with multistability
show various features of interwovenness in 
basins of attraction, such as
fractal \cite{f1, f2, fr, fn}, riddled \cite{fr, r1, r2, rn, rs}
and intermingled \cite{i1}  basin boundaries.
Recently they became objects of
intense investigations with related  
subjects including synchronization \cite{rs, s1}
and noise effects \cite{fn, rn, n1}.
 
In this paper, geometrical regularities of 
basin structures for two-dimensional simply doubled logistic map
at the edge of chaos is investigated. 
These basin structures are found to have a regularity
which shows global interwovenness and overall interrelation.

Simply doubled logistic map is
\begin{equation}
\left\{
\begin{array}{c}
 x_{t+1} \\
 y_{t+1}
\end{array}
\right\} =
\left\{
\begin{array}{c}
\mu x_t (1-x_t )\\
\mu y_t (1-y_t )
\end{array}
\right\}.
\label{simple}
\end{equation}
It is a very simple two-dimensional map
which is physically trivial.
Nevertheless, it shows very regular
interwovenness which may have referential
significances for other interwoven structures.

\begin{figure}[!]
\includegraphics[width=6cm]{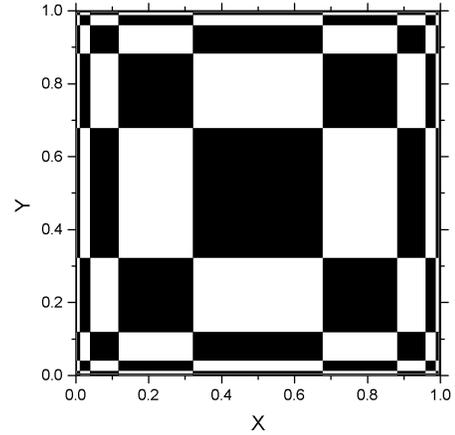}
\caption[0]{Two basins of attractions at $\mu=3.4$.
Dark regions are the basin of the``in-phase'' attractor
while white ones are for the ``out-of-phase'' attractor.}
\label{p-2}
\end{figure}

In the period-2 region, the map has two attractors 
``in-phase'', $(x_1, x_1) \leftrightarrow (x_2, x_2)$ 
 and ``out-of-phase'',
$(x_1, x_2) \leftrightarrow (x_2, x_1)$ 
where $x_1$ and $x_2$ are cyclic
points of the logistic map,
$x_{t+1}=F_{\mu}(x_{t})=\mu x_{t}(1-x_{t})$.
Likewise, the Eq. (\ref{simple}) has $2^p$ attractors 
in the $2^p$-cycle region
with phases,
\begin{equation}
{y_t} = {x_{t+m}}
\quad for
\quad m=0,1,2, \cdots, 2^p-1. \nonumber
\label{phase}
\end{equation}
With $\mu=3.4$ in  period-2 region,
the map has two basins of attraction as shown
in Fig. \ref{p-2}, where dark regions are the basin of 
the ``in-phase'' attractor while white rectangles
are that of the ``out-of-phase'' attractor.
\begin{figure}[!]
\includegraphics[width=6cm]{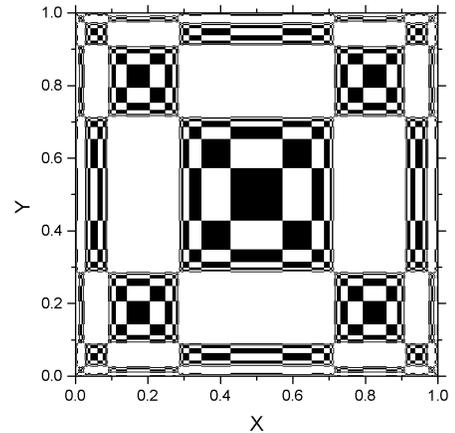}
\caption[0]{The basin of ``in-phase'' attractor (dark regions)
with $\mu=3.54$(period-4). Basins of the other three attractors
will occupy white regions.
}
\label{p-4}
\end{figure}
\begin{figure}[!]
\includegraphics[width=6cm]{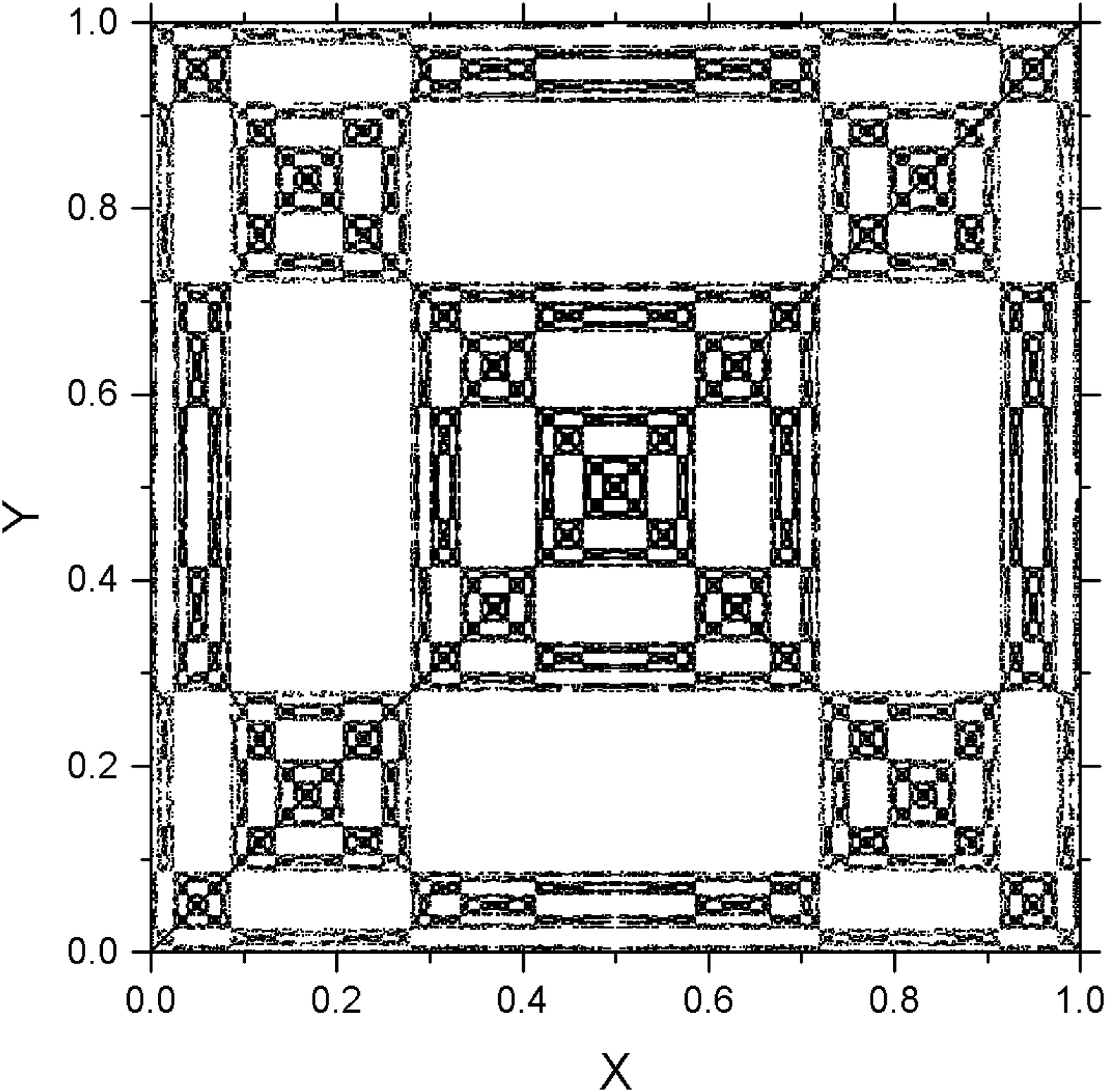}
\caption[0]{The basin of the ``in-phase'' attractor
with $\mu=3.57$ very near to the edge of chaos.
}
\label{b0}
\end{figure} \begin{figure}[!]
\includegraphics[width=6cm]{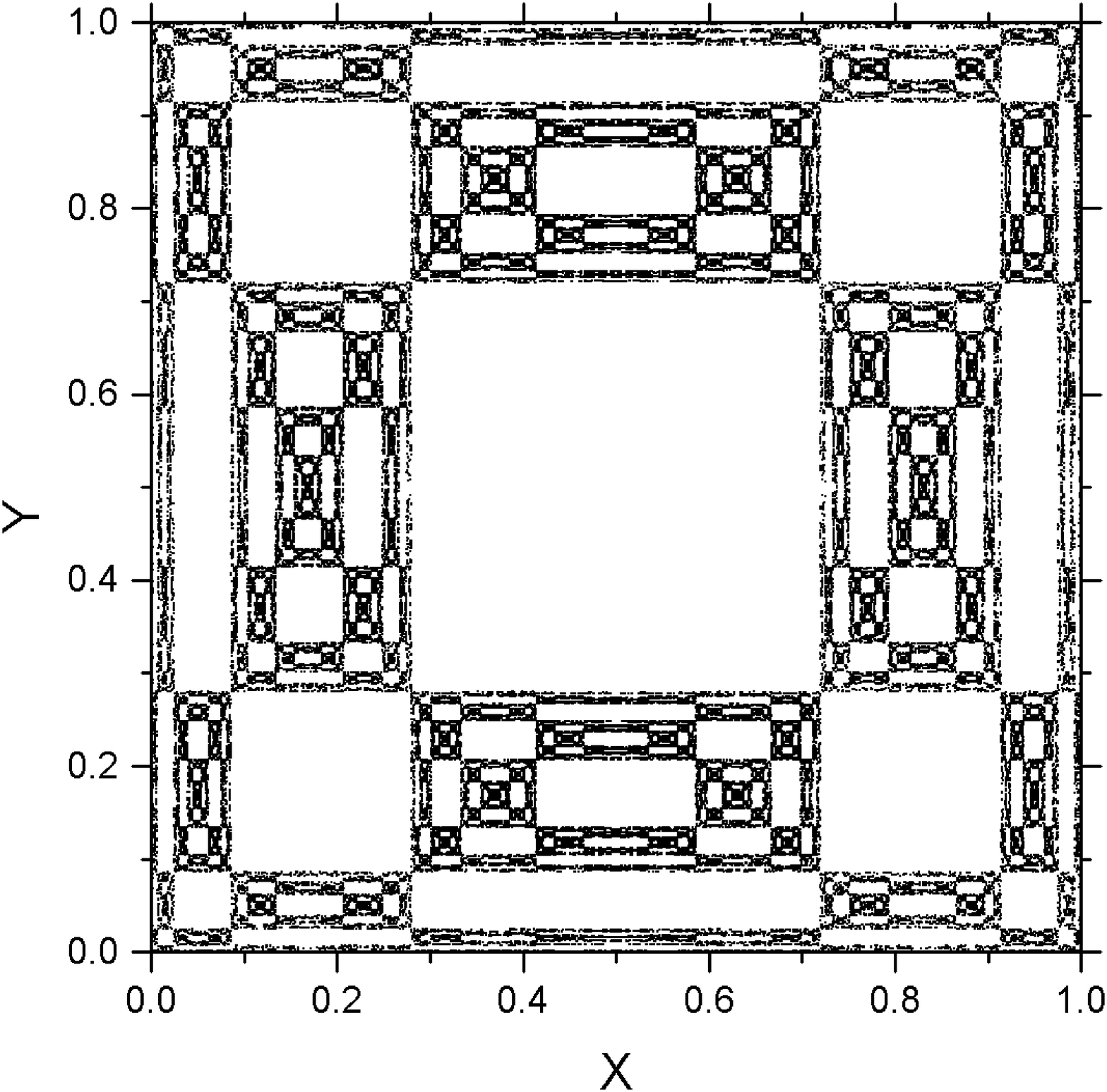}
\caption[0]{The basin of an attractor with phase
$y_t  =x_{t+1}$.
} \label{b1}
\end{figure}
\begin{figure}[!]
\includegraphics[width=6cm]{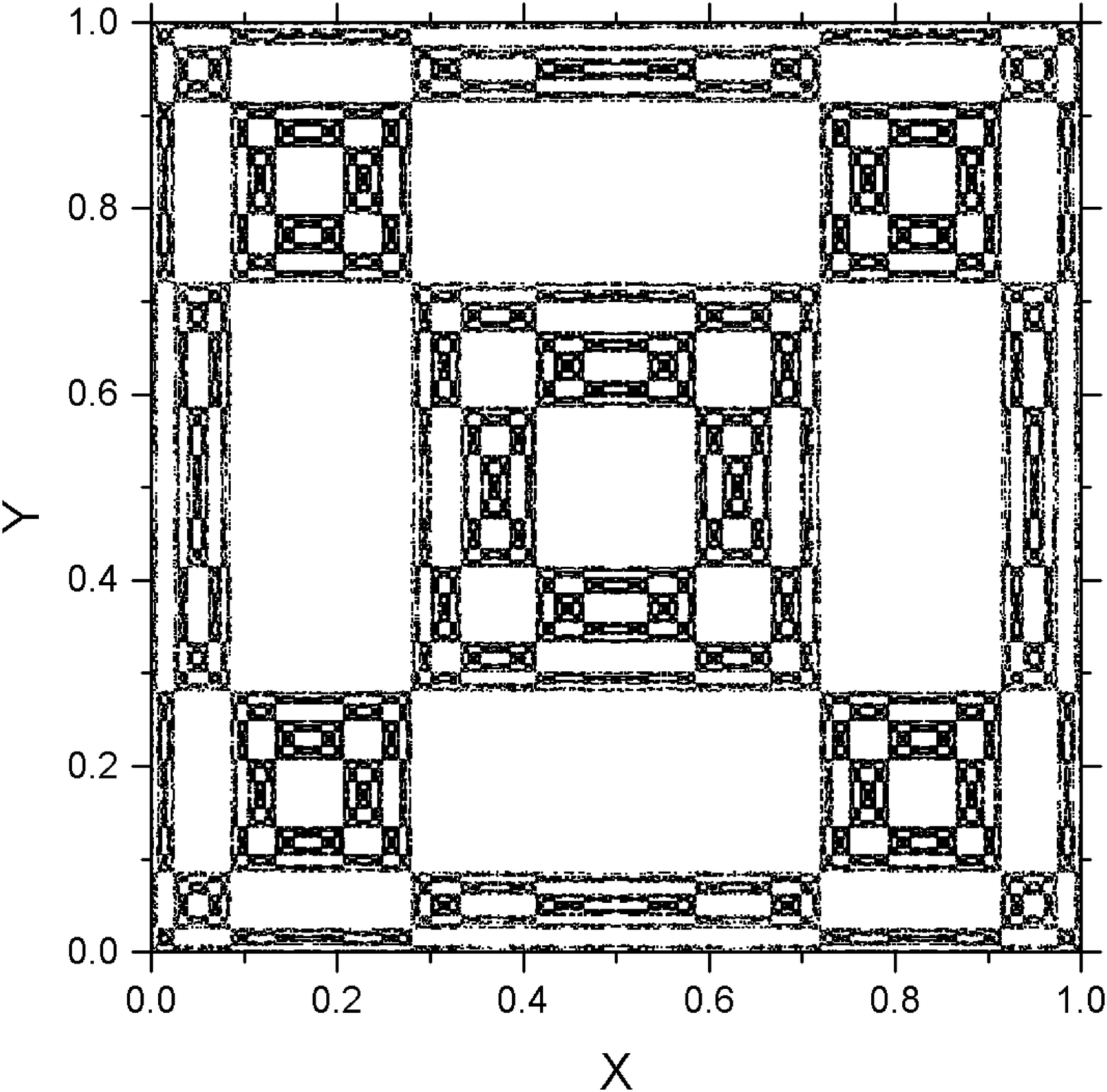}
\caption[0]{The basin of an attractor with phase
$y_t  =x_{t+2}$. 
}
\label{b2}
\end{figure}
\begin{figure}[!]
\includegraphics[width=6cm]{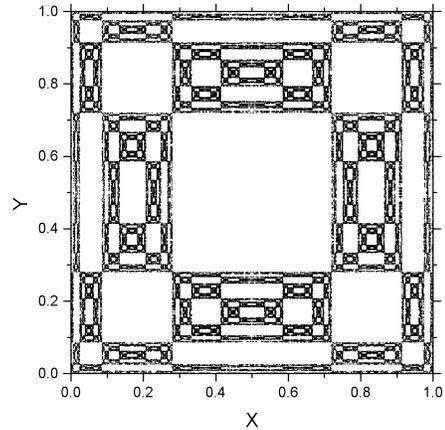}
\caption[0]{The basin of an attractor with phase
$y_t  =x_{t+3}$. }
\label{b3}
\end{figure}

Divisions of these rectangular regions can be given with
straight lines, $x=x_n$, $x=1-x_n$, $y=x_n$,
and $y=1-x_n$ with $x_n$'s as
\begin{eqnarray}
{x_n} =\max\{x|F^{(n-1)} _{\mu}(x)=F^{(n)} _{\mu}(x)\}  \nonumber \\
for  ~ n=1,2,3,\cdots. 
\label{xn}
\end{eqnarray}
The basin of the
``in-phase'' attractor with $\mu=3.54$ in the period-4 region  
is shown in Fig. \ref{p-4} with dark regions,
while white regions will be occupied by 
the other three basins. 
For the convenience of
later discussions, let us designate
rectangles white or dark in Fig. \ref{p-2} as ``primary'', and
inner rectangles in ``primary'' ones as seen in Fig. \ref{p-4} 
as ``secondary'' rectangles.
Actually there are an infinite number of ``primary''
rectangles with a wide range of sizes and proportions in Fig. \ref{p-2}.
Consequently, most of them have infinitesimal area. 

Our main concern is basin structures with 
$\mu=\mu_\infty$, on the edge of chaos
where there exist an infinite number of attractors.
Among them,
a few basins of attractors with small phase differences
between $x$ and $y$ are selected and presented  in
Figs. \ref{b0} - \ref{b3} ($y_t  =x_t $, 
 $y_t  =x_{t+1}$,  $y_t  =x_{t+2}$,  $y_t  =x_{t+3}$ in sequence). 
\begin{figure}[!]
\includegraphics[width=6cm]{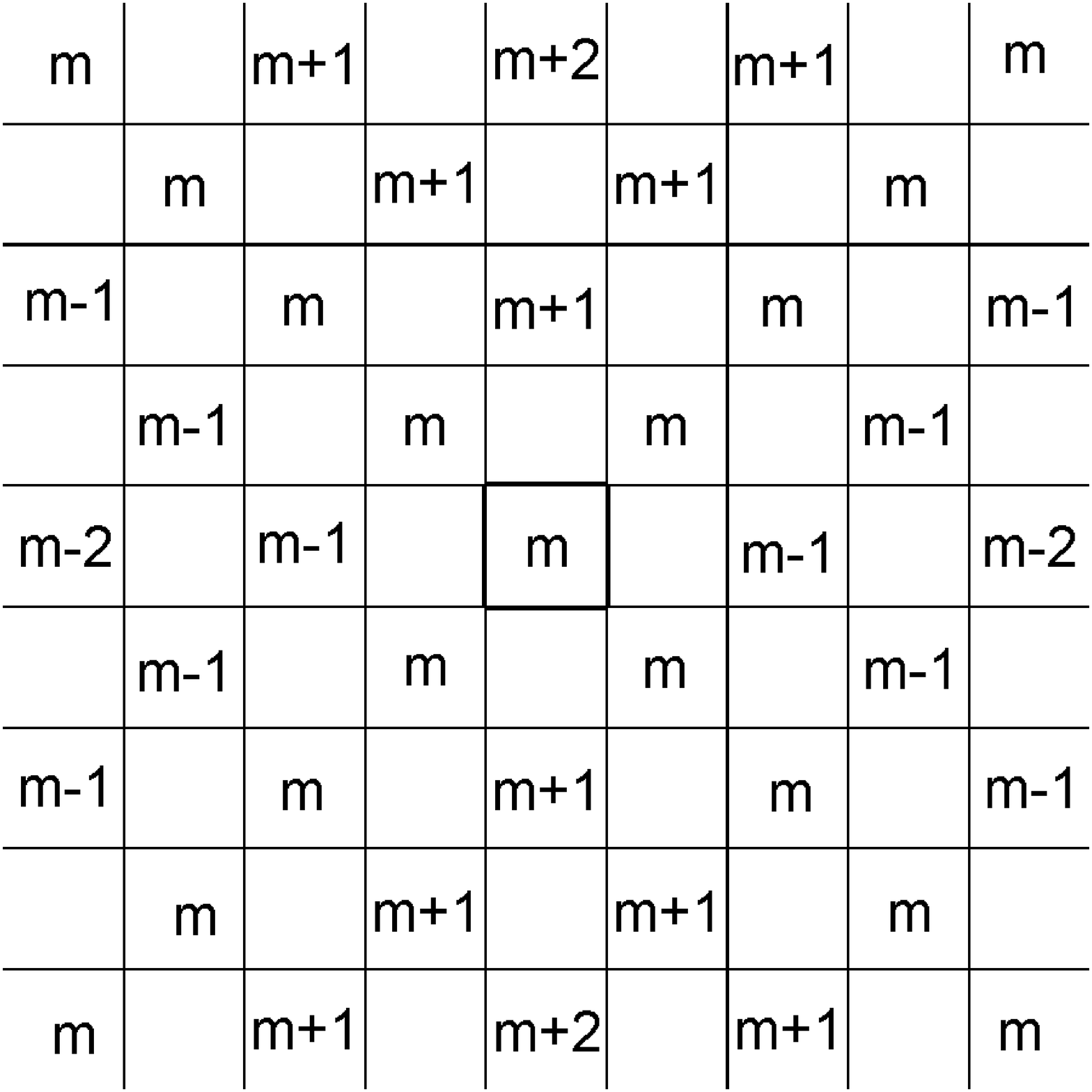}
\caption[0]{Global pattern of basin for an attractor with
phase $y_t =x_{t+2m}$:$[2m]$. Squares denotes
``primary'' rectangles which actually do not have equal areal sizes
and proportions.
The cental ``primary'' square is
occupied by $[m]$.
}
\label{2m}
\end{figure}
\begin{figure}[!]
\includegraphics[width=6cm]{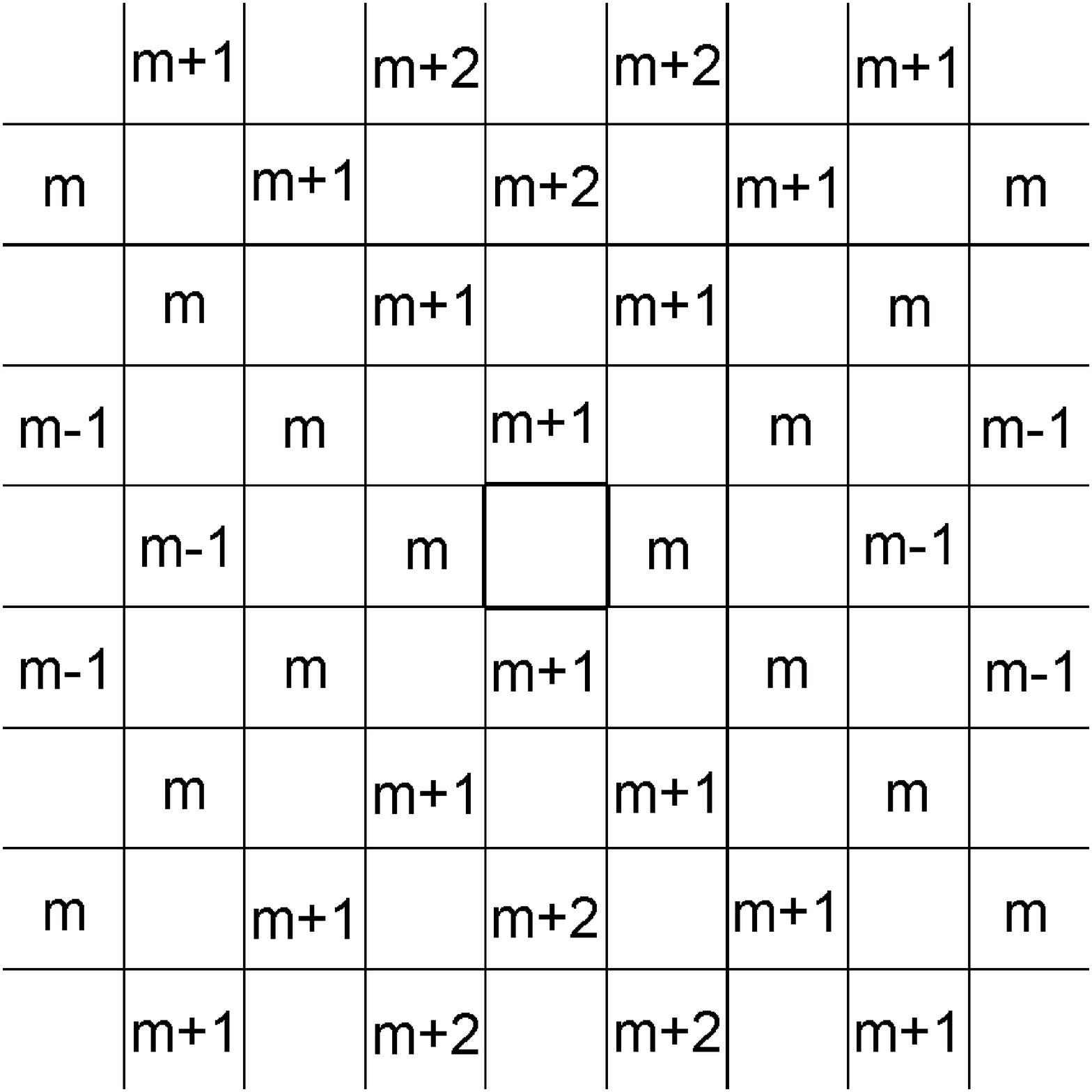}
\caption[0]{Structure of basin with
phase $y_t =x_{t+(2m+1)}$:$[2m+1]$.
The central square is vacant and the nearest rectangles are
occupied by $[m]$ and $[m+1]$. }
\label{2m+1}
\end{figure}
In these Figures, we can see 
repeated appearances of some patterns
in various sizes
with winding curves,
thus, one might suspect the existence of
a carpeting regularity ruling all
basin structures.
We may ask 
how the patterns of those basin carpets 
are composed,
and what those curves are.
 
\begin{figure}[!]
\includegraphics[width=6cm]{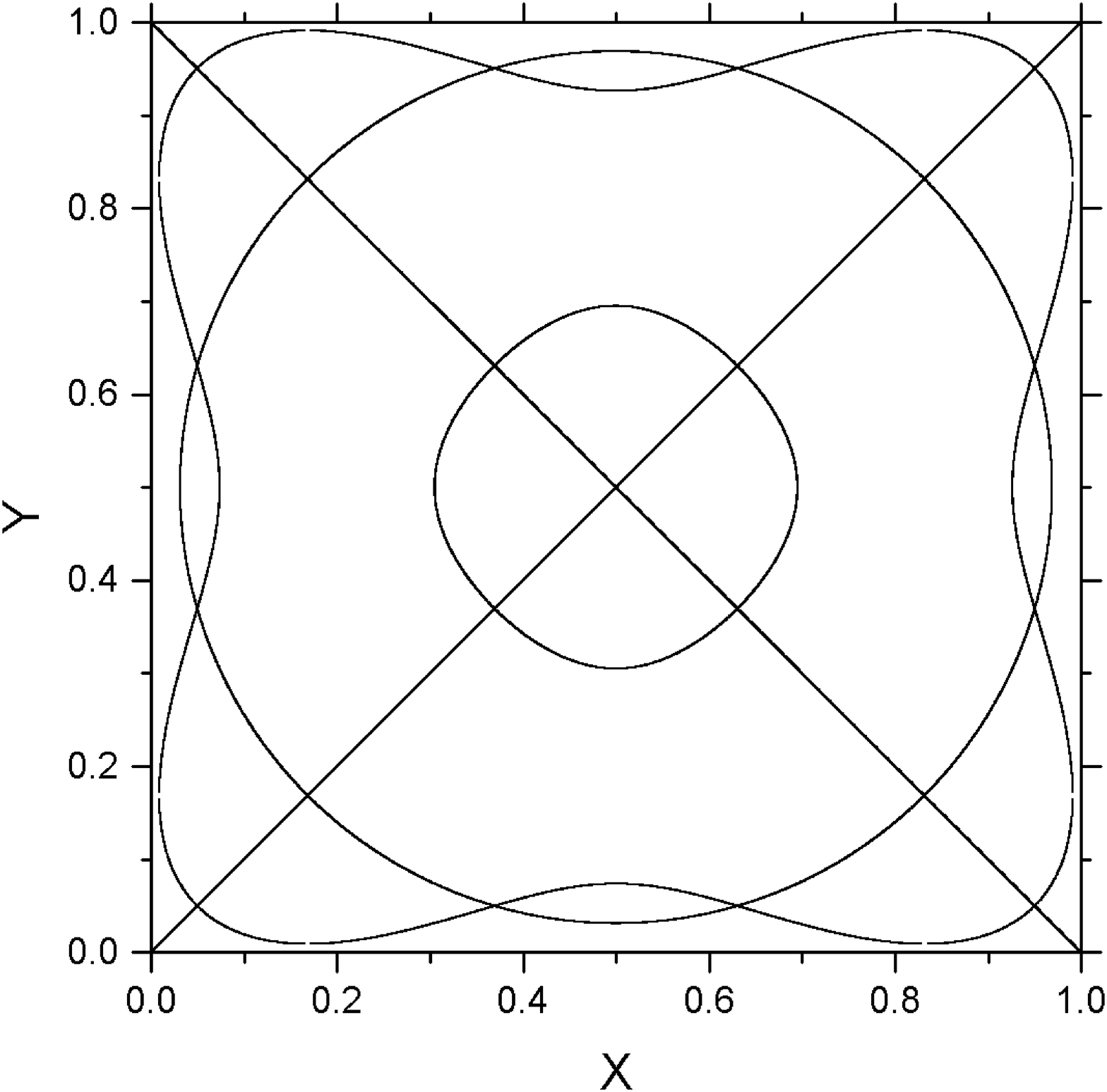}
\caption[0]{Curves appearing in the Fig. \ref{b0}.
The outermost circle is expressed by Eq. (\ref{circle})
and Eq. (\ref{loop1}) is the expression for the winding loop 
and inner circlelike loop. 
}
\label{c0}
\end{figure}

A simple and effective way to answer the first question
is to elucidate the 
overall interrelation among basin structures
rather than enumerate each of them.
Examining every ``primary'' rectangle
in global basin structures,
its pattern is found to have correspondence with
a global basin structure.
It is needed to magnify ``primary'' rectangles whose sizes
are too small for direct examination.
The result is summarized in Fig. \ref{2m} and Fig. \ref{2m+1} 
which represent central parts of $[2m]$ and $[2m+1]$,
global basin structures of two attractors with phases  
$y_t =x_{t+2m}$ and $y_t =x_{t+(2m+1)}$.
Here we let a symbol $[m]$ denote the
pattern of the global basin structure for an attractor 
with phase $y_t =x_{t+m}$, and
the letter $m$ in Fig. \ref{2m} and Fig. \ref{2m+1} has 
the same meaning.
We should remark that 
[m] represents only the pattern of the basin structure 
and not its exact proportion or size,
and all ``primary'' rectangles are presented as squares
in Fig. \ref{2m} and Fig. \ref{2m+1} 
in spite of differences in area and proportion.

The rectangles are occupied in a checker pattern and
the largest central ``primary'' square as in Figs. \ref{b0} - \ref{b3} 
is oriented at the center of
Fig. \ref{2m} or Fig. \ref{2m+1}.
In $[2m]$ (Fig. \ref{2m}), the central square 
and those along the diagonals 
are occupied by $[m]$, 
this is followed by $[m-1]$, $[m-2]$, etc. 
moving horizontally away from the
diagonals and
by $[m+1]$, $[m+2]$, etc. moving vertically away from the
diagonals. 
For $[2m+1]$, the central square is empty
and the nearest rectangles are occupied with $[m]$ and $[m+1]$ 
as shown in Fig. \ref{2m+1}. 
Accordingly, global patterns of Figs. \ref{b0} - \ref{b3}
can be symbolized as $[0]$, $[1]$, $[2]$ and $[3]$ respectively.
The same rule will hold for the relation between ``primary'' 
rectangles and ``secondary'' ones and also for rectangles of further levels.

The pattern of basin carpet $[m]$ can be defined 
by $\{ \cdots{b_3}{b_2}{b_1}\}$, the binary notation of 
an integer number $m$,
as the last digit $b_1$ determines occupations of ``primary'' rectangles
and $b_2$ arranges ``secondary'' ones.
With their binary variations, these basin patterns remind us of the
T'ai Chi (Tai Ji) diagram in I jing (Yijing, Book of Changes) which is 
the central frame in chinese cosmology \cite{t1}.

We can see the emergence of a
large number of curves which are crossing each other 
in Figs. \ref{b0} - \ref{b3}.
The appearance of curves seems to be
a consequence of the edge of chaos.
In Fig. \ref{b0}, noticeable ones are
diagonal lines, a circular shape and
more winding curves.
We attempt to express them respectively as 

\begin{eqnarray}
\bar{y}&=&\bar{x},  \nonumber \\  
\bar{y}&=&-\bar{x}, \nonumber 
\end{eqnarray}
\begin{equation}
 \bar{x}^2 +\bar{y}^2 +A_1=0,
\label{circle}
\end{equation}
and
\begin{equation}
\bar{x}^4+\bar{y}^4+B_1 \bar{x}^2
+B_1 \bar{y}^2+B_2 =0, 
\label{loop1}
\end{equation}
with proper constants which can be expressed with
$x_n$'s of Eq. (\ref{xn}) as
\begin{eqnarray}
A_1&=&B_1=- \bar{x}_1^2 - \bar{x}_2^2, \nonumber \\  
B_2&=&\bar{x}_1^2 \bar{x}_2^2
  +\bar{x}_1 ^2\bar{x}_3 ^2
  +\bar{x}_2 ^2\bar{x}_3 ^2
  - \bar{x}_3^4. \nonumber  
\end{eqnarray}
Here we used $\bar{x}$, $\bar{y}$ and ${\bar{x}_n}$
as $\bar{x}\equiv x-\frac{1}{2}$, 
$\bar{y}\equiv y-\frac{1}{2}$ and $\bar{x}_n\equiv x_n-\frac{1}{2}$
for simplicity of expressions.
Curves of these expressions are depicted in Fig. \ref{c0},
they show exact coincidences with the curves in Fig. \ref{b0}.

In this paper,
we have studied orderly interwoven patterns in basins
of double logistic map.
Interwovenness is beautiful 
from an aesthetic viewpoint and certainly has
some geometrical regularities.
But for the most part they are emergent and mystic, so,
direct apprehension of them is not easy.
We have dealt with interwoven patterns
which are exceptionally regular, with a definite global
order.
This regularity may have referential significance
for other interwoven structures,
as well as for basin structures of coupled maps.
The order in these patterns may be visible facets of 
an exquisite regularity for which there 
remains much room for further investigation.
We see that these basin structures have impressive interrelations
which might be called  \emph{all in one and one in all},
all the basin structures are contained in a
basin which takes part in building up the all {\it vice versa}
over a wide range of scale.
\emph{This interrelation
suggests an interesting idea to us about the wholeness
and the elementariness.} 
Interwovenness may be ubiquitous, 
and is an interesting aspect
of nonlinear phenomena in nature. 
Clarification of  the regularity
might be a way to understand
the nonlinear features.

\begin{acknowledgements}
The author acknowledges a critical reading of the manuscript and
also fruitful discussions with Michael Walker.
\end{acknowledgements}

\end{document}